# Intercalation effect on hyperfine parameters of Fe in FeSe superconductor with $T_c$ = 42 K

Sergii I. Shylin[1,2], Vadim Ksenofontov[1], Stefan J. Sedlmaier[3], Simon J. Clarke[3], Simon J. Cassidy,[3] Gerhard Wortmann[4], Sergey A. Medvedev[5] and Claudia Felser[5]

[1] *Institute of Inorganic and Analytical Chemistry, Johannes Gutenberg-University Mainz, Staudingerweg 9, D-55099 Mainz, Germany*
[2] *Department of Chemistry, Taras Shevchenko National University of Kyiv, Volodymyrska 64/13, 01601 Kyiv, Ukraine*
[3] *Department of Chemistry, University of Oxford, Oxford OX1 3QR, UK*
[4] *Department of Physics, University of Paderborn, D-33095 Paderborn, Germany*
[5] *Max Planck Institute for Chemical Physics of Solids, D-01187 Dresden, Germany*

PACS `74.70.Xa` – Pnictides and chalcogenides
PACS `76.80.+y` – Mössbauer effect; other γ-ray spectroscopy
PACS `74.62.Bf` – Effects of material synthesis, crystal structure, and chemical composition

**Abstract** – $^{57}$Fe-Mössbauer spectra of superconducting $\beta$-FeSe, the Li/NH$_3$ intercalate product and a subsequent sample of this intercalate treated with moist He gas have been measured in temperature range 4.7 – 290 K. A correlation is established between hyperfine parameters and critical temperature $T_c$ in these phases. A strong increase of isomer shift upon intercalation is explained by a charge transfer from the Li/NH$_3$ intercalate to the FeSe layers resulting in an increase of $T_c$ up to 42 K. A significant decrease of the quadrupole splitting above 240 K has been attributed to diffusive motion of Li$^+$ ions within the interlamellar space.

**Introduction.** – The recent development of iron-based superconductors has prompted extensive research focusing on the new synthetic approaches to high-$T_c$ systems and on studying the superconducting pairing mechanism in these materials. Particular attention has been paid to iron chalcogenide compounds, notably the tetragonal polymorph of FeSe ($\beta$-FeSe ($T_c$ = 8 K)) [1, 2], where a marked increase of $T_c$ up to 37 K has been observed under pressure [3]. Increase of $T_c$ in FeSe related compounds has also been obtained in a variety of its derivatives with cationic spacers between the FeSe layers consisting of alkali metal ions with the general formula A$_x$Fe$_{2-y}$Se$_2$ (A = K$^+$ ($T_c$ = 31 K) [4], Rb$^+$ ($T_c$ = 32 K) [5], Cs$^+$ ($T_c$ = 27 K) [6]). However these compounds exhibit a complex phase separation, where only minority FeSe-like regions of the bulk samples are responsible for superconductivity [7-10]. For the basic FeSe system, also partial substitution Se by Te results in an increase of T$_c$ up to 13 K in FeSe$_{0.5}$Te$_{0.5}$, which is attributed mainly to chemical pressure [11].

Further alternative synthesis routes of FeSe-based superconductors have been undertaken. Due to solubility of alkali, alkaline-earth metals as well as Eu and Yb in liquid ammonia and some amines [12], one of the approaches is the intercalation reaction of FeSe under these conditions resulting in products accommodating electropositive metal ions, ammonia and amide ions between the FeSe layers. Thus, successful intercalation reactions with Li, Na, K, Ca, Sr, Ba, Yb and Eu with FeSe by the ammonothermal method at temperatures around -78 °C have been reported [13]. $T_c$ values up to 46 K in these compounds have been found and recent computational investigations suggest that the enhancement of $T_c$ relative to that in the FeSe parent material is a consequence of making the Fermi surface more two-dimensional and of electron doping [14]. Superconductors have also been obtained by the intercalation of alkali metals along with other molecules, notably pyridine [15], although these compounds are not yet characterised in detail. Following the ammonothermal intercalation approach, Burrard-Lucas *et al*. have characterised Li$_x$(NH$_2$)$_y$(NH$_3$)$_{1-y}$Fe$_2$Se$_2$ (x=0.6; y=0.2) containing Li$^+$ and amide ions as well as ammonia molecules acting as the spacer layer between FeSe layers, which exhibits bulk superconductivity (superconducting volume fractions of about 50%) below 43 K [16], and Sedlmaier *et al*. have identified a more ammonia-rich intercalate by probing the intercalation reaction *in situ* [17]. Using neutron powder diffraction, the crystal structure of the deuterated intercalated compounds has been unambiguously determined. It showed evidence for weak N–D···Se hydrogen bonds and the refined compositions revealed that the amide (ND$_2^-$) content was lower than the Li$^+$ content indicating an electron transfer to the FeSe layers and, therefore, into the conducting band responsible for superconductivity. Although the superconducting mechanism in iron-based superconductors still remains enigmatic, it has been proposed that magnetic fluctuations could play a role in the pairing mechanism of superconducting charge carries [18]. Apparently, both spin fluctuations and the density of states at the Fermi level are among the major factors effecting $T_c$. Taking into account that the conduction band at the Fermi level is dominantly formed by Fe-$3d$ orbitals, Mössbauer spectroscopy can be a useful tool to study Fe-based superconductors.

Here we report the detailed temperature-dependent $^{57}$Fe-Mössbauer study of non-intercalated $\beta$-FeSe (**1**), the Li/ammonia intercalate product (**2**) and a subsequent sample of this intercalate treated with moist He gas (**3**). Our observations demonstrate that $d$-electron density on Fe ions increases upon intercalation and decreases after subsequent treatment with water vapour along with $T_c$. Moreover, the Mössbauer data for Li/ammonia intercalate indicate the diffusive motion of Li$^+$ ions within the interlayer space.

**Experimental details.** – The sample of tetragonal $\beta$-FeSe (**1**) was synthesized by heating a stoichiometric mixture of iron powder (Johnson-Matthey, 99.98%) and selenium shots (ALFA 99.99%) and structurally characterized as described elsewhere [2]. $\beta$-FeSe used as the precursor to **2** was synthesised in a similar way from iron powder (ALFA 99.995%) and selenium shots (ALFA 99.99%).

The intercalate Li$_x$(NH$_2$)$_y$(NH$_3$)$_{1-y}$Fe$_2$Se$_2$ (**2**) was synthesised from $\beta$-FeSe, Li metal (99%, Aldrich) and ammonia (99.98%, BOC) [16]. All manipulations of solids were performed under argon. $\beta$-FeSe and Li (molar ratio 2:1) were placed in a Schlenk tube and this, along with a cylinder of ammonia was connected to a Schlenk line. After evacuation of the system and cooling the Schlenk tube down to 195 K (dry ice/isopropanol bath), the valve on the ammonia cylinder was opened allowing ammonia to condense onto the reactants. The mixture was stirred for 30 min, and then the Schlenk tube was allowed to warm to room temperature enabling excess ammonia to evaporate *via* a mercury bubbler. After brief evacuation the powder of **2** was isolated as a black powdery solid in an argon-filled glove box.

After the Mössbauer measurements of **2** had been performed under a dry He atmosphere, **2** was partially hydrolysed in the closed sample volume of the Mössbauer cryostat by introducing moist helium gas into the sample space. After 24 hours exposure to this moist helium atmosphere, the exchange gas was replaced and the modified sample (**3**) was kept under dry helium during subsequent measurements.

Purity of the samples was confirmed using a Philips PW1730 X-ray diffractometer (CuK$\alpha_1$/K$\alpha_2$ radiation). Magnetic susceptibility measurements were performed using a Quantum Design SQUID magnetometer in the field of 20 Oe in the temperature range of 2–55 K using zero-field-cooled (ZFC) and field-cooled (FC) measurements. $^{57}$Fe-Mössbauer spectra were recorded in transmission geometry with a $^{57}$Co source in a rhodium matrix using a conventional constant-acceleration Mössbauer spectrometer equipped with a nitrogen/helium bath cryostat in the temperature range of 4.7–290 K. Isomer shifts are given relatively to an α-Fe foil at ambient temperature. Fits of the experimental Mössbauer data were performed using the Recoil software [19]. Hyperfine parameters uncertainties given in parentheses were evaluated using the covariance matrix of the fit. The absorbers were prepared by placing the powdered samples (around 30 mg) in plastic holders. All the sample preparation procedures were performed in an argon glove box with an O$_2$ and H$_2$O content below 0.5 ppm.

**Results and discussion.** – X-Ray powder diffraction (XRPD) patterns for **2** and **3** are shown in Fig. 1. The peaks observed for **2** are accounted for the Li$_x$(NH$_2$)$_y$(NH$_3$)$_{1-y}$Fe$_2$Se$_2$ superconductor phase with a structure based on that of ThCr$_2$Si$_2$ with a body centred tetragonal unit cell ($a$ = 3.785(2) Å, $c$ = 16.914(9) Å) [16]. The peak shapes of this sample suggest some imperfection in the stacking of the layers. On exposure to moist helium to obtain **3**, there is evidence for partial deintercalation. However, the majority phase may be accounted for using the model-independent Pawley method on a body centred tetragonal cell ($a$ = 3.8572(2) Å, $c$ = 16.011(2) Å) with a $c$ lattice parameter greatly reduced as compared with sample **2**. A minor part of the sample decomposed to a phase resembling $\beta$-FeSe but with much broader Bragg peaks. Further weak reflections in the pattern of **3** could not readily be identified. The dominant phase in the powder pattern of **3** suggests that another intercalate of FeSe is produced by the gentle hydrolysis treatment.

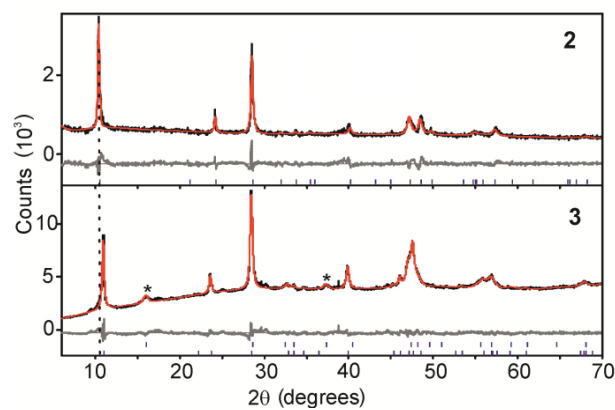

Fig. 1: (Colour on-line) X-Ray powder diffraction patterns of **2** and **3**. The data (black line), Pawley-type fit (red line) and difference (grey line) are shown. Tick marks are for the main body centred tetragonal phase (lower sets) and the FeSe-like phase in **3** (upper set). The dotted line is used to show the shift in the 002 reflection after mild hydrolysis. The asterisks indicate reflections arising from a $\beta$-FeSe-like phase.

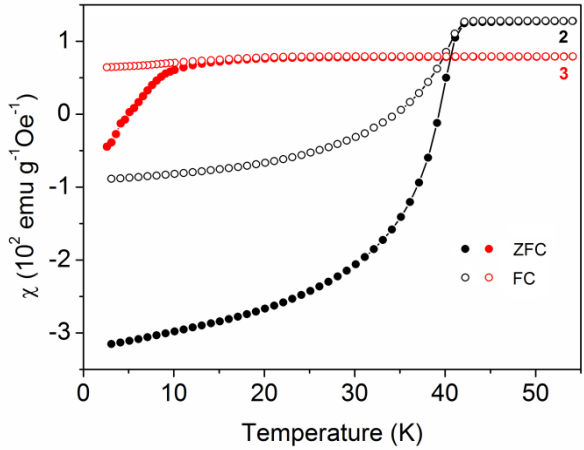

Fig. 2: (Colour on-line) Magnetic susceptibility measurements on **2** (black) and **3** (red) in the ZFC-FC mode.

Magnetic measurements of **1**–**3** show superconductivity in all samples. In agreement with previous studies, the non-intercalated compound (**1**) exhibits a superconducting transition at 8 K [1, 2] and the intercalated one (**2**) has a transition at 42 K (Fig. 2) [16]. After exposure to a moist helium atmosphere the superconducting properties of the intercalated sample change dramatically. From the susceptibility measurements, the onset transition temperature for **3** is determined to be 12 K, and diamagnetic volume fraction noticeably decreases below $T_c$ in comparison with **2** (Fig. 2).

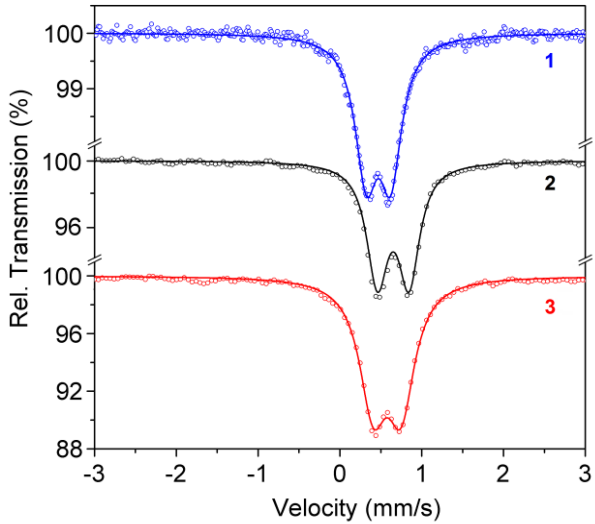

Fig. 3: (Colour on-line) Mössbauer spectra of **1** (blue), **2** (black) and **3** (red) recorded at 100 K in the ± 3 mm/s velocity range.

Mössbauer spectra of **1**–**3** recorded at 100 K are shown in Fig. 3 and the derived hyperfine parameters (isomer shift δ, quadrupole splitting $\Delta E_Q$ and line width Γ) are summarized in Table 1. All spectra consist of single paramagnetic doublets having close values of linewidth that evidence one single Fe site in the structures. $\Delta E_Q$ of **1** reflects a tetragonal structure of the parent compound and both $\Delta E_Q$ and δ indicate a low-spin (LS) state of divalent iron [2]. The parameters of its derivatives, **2** and **3**, indicate changes in electron density on the Fe nuclei (δ) as well as in the local distortion symmetry ($\Delta E_Q$). Assuming that the FeSe layers remain intact and the first coordination sphere of Fe does not change after the intercalation procedure and after the mild hydrolysis to produce **3**, such a difference in δ means an increase in *d*-electron density. Indeed, taking into account the composition of **2**, $Li_x(NH_2)_y(NH_3)_{1-y}Fe_2Se_2$ (x=0.6; y=0.2), one can calculate the formal oxidation state of Fe to be +1.8, whilst in **1** it is +2. Apparently, the Mössbauer measurements together with the above XRPD data show that the reaction to convert **2** into **3** does not simply convert **2** back to pure FeSe. This suggests that a range of further intercalates of FeSe may exist, and may merit further investigation.

Table 1: Hyperfine parameters of **1**–**3** at 100 K.

| Sample | δ (mm/s) | $\Delta E_Q$ (mm/s) | Γ (mm/s) |
|---|---|---|---|
| **1** | 0.557(2) | 0.286(3) | 0.181(4) |
| **2** | 0.641(2) | 0.396(4) | 0.178(4) |
| **3** | 0.579(3) | 0.337(6) | 0.179(5) |

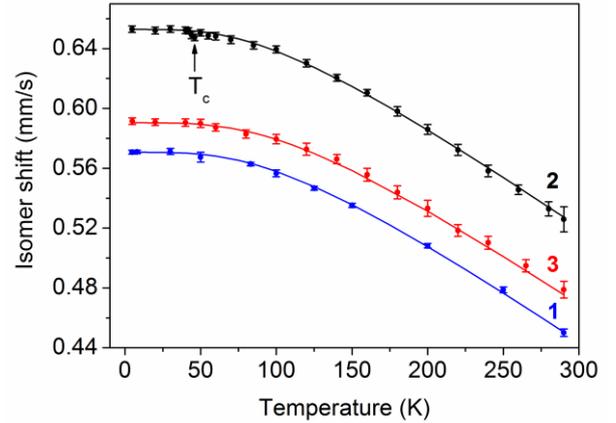

Fig. 4: (Colour on-line) Temperature dependences of the isomer shift for **1** (blue), **2** (black) and **3** (red). The fitting curves were adjusted with the Debye model (eq. 1).

Additional information on the properties of the present samples was obtained from Mössbauer spectra recorded in the temperature range 4.7 – 290 K. Variation of δ with temperature is presented in Fig. 4. The observed decrease of δ with increasing temperature is caused by the second-order Doppler shift, used here to derive Debye temperatures ($\Theta_D$) by fitting the temperature dependence of δ using the Debye model [20],

$$\delta(T) = \delta(0) - \frac{9k_B T^4}{2\Theta_D^3 Mc} \int_0^{\Theta_D/T} \frac{x^3 dx}{e^x - 1}, \qquad (1)$$

providing $\Theta_D$ = 390(5) K, 365(5) K and 420(5) K for **1**, **2** and **3** respectively. Determination of $\Theta_D$ using eq. 1 results often in values, which are higher than $\Theta_D$ obtained from density of phonon states (DOS), caused by other factors, *e.g.* thermal expansion. In the present case, we derived for sample **1** a lower value, $\Theta_D$ = 285 K, in our study of the local DOS using nuclear inelastic scattering (NIS) [21]. Comparison of the Debye temperatures of **1**–**3** indicates some softening of the phonon spectrum after intercalation to produce **2** and its hardening after subsequent mild hydrolysis to produce **3**. A similar softening of the phonon spectrum upon intercalation of FeSe has been also observed in our recent NIS measurements [22].

A marked increase of $\Delta E_Q$ by about 40% between **1** and **2** reflects a significant increase of the electric field gradient (EFG) upon intercalation. This finding is in accordance with the crystal structure [16] indicating a larger distortion of FeSe$_4$ tetrahedra in **2**, as expressed by the variance of the polyhedral angles $\sigma_\Theta^2$ = 22.49 and 32.56 for **1** and **2**, respectively. However, such an increase of $\Delta E_Q$ might be also caused by other reasons as evidenced by the Mössbauer data at different temperatures. Dependences of $\Delta E_Q$ on temperature are shown in Fig. 5. A slight increase in $\Delta E_Q$ of **1** upon cooling is typical for LS Fe(II) tetrahedral compounds and can be plausibly fitted using the simplified model for tetragonal distortion in an axial electric field [23]:

$$\Delta E_Q(T) = \Delta E_Q(0) \cdot \frac{1-e^{-E_0/k_BT}}{1+2e^{-E_0/k_BT}}, \qquad (2)$$

where $E_0$ is energy of a crystal field splitting. For **1**, a splitting value of 0.072(2) eV has been found to increase up to 0.078(2) eV after intercalation (**2**) and to 0.089(2) eV after hydrolysis treatment (**3**).

Although this model is completely appropriate and describes well the temperature dependence of $\Delta E_Q$ for **1** in the whole temperature range, an anomalous decrease of $\Delta E_Q$ above 240 K for **2** and **3** is observed. We suppose that such behaviour of $\Delta E_Q$ *vs.* T curves is associated with the motion of Li$^+$ ions. It is known that in certain cases Li-containing solids may show a mobility of Li$^+$ that usually is monitored by the NMR technique [24]. Herein, we observe an essential decrease of $\Delta E_Q$ above 240 K upon heating the Li-intercalated specimens **2** and **3**. Cooling then the samples, we notice that the behaviour of $\Delta E_Q$ in the region of 240–290 K is completely reversible and does not show thermal hysteresis. At the same time, $\delta$ (Fig. 4), as well as the line width, does not show any peculiarities in this temperature region. It allows concluding that this behaviour of $\Delta E_Q$ is not caused by a structural transition or other reasons affecting the electron density on Fe nuclei. Therefore, we attribute this strong decrease of $\Delta E_Q$ above 240 K to the thermal activation of the Li$^+$ motion. Indeed, according to the crystal structure of **2** obtained from neutron powder diffraction measurements at 293 K, Li$^+$ ions are distributed over two sites (2*b*, 4*c*, *I4/mmm*) separated by less than 2 Å, each with fractional occupancies of ~0.25 [16]. The large displacement ellipsoids for these ions are consistent with rapid hopping between sites at room temperature. We assume that at low temperature the Li$^+$ ions are arranged in an ordered array on these sites, similar as described elsewhere [25], thereby providing a defined contribution to the EFG at the Fe sites. Above 240 K the thermoactivated motion of the Li$^+$ ions between their lattice sites becomes rapid enough on the timescale of the Mössbauer experiment of $10^{-8}$ s to lead to an averaging of the electric field gradient caused by Li$^+$ ions and their contribution to the observed $\Delta E_Q$ decreases. Assuming that the Li-induced EFG fluctuates between two opposite directions, a theoretical model [26] for a fluctuating EFG has been applied to derive the activation energy E$_a$ for the Li$^+$ hopping from the variation of the EFG in the region of 240–290 K:

$$\Delta E_Q^{Li}(T) = \Delta E_Q(T) \cdot \frac{1-e^{-E_a/k_BT}}{1+e^{-E_a/k_BT}}, \qquad (3)$$

where $\Delta E_Q^{Li}$(T) describes the variation of the quadrupole splitting considering the Li$^+$ motion (dashed lines in Fig. 5), and $\Delta E_Q$(T) are the values obtained by eq. 2 above 240 K (solid line in Fig. 5). The E$_a$ values for **2** and **3** have been determined as 0.050(14) and 0.058(7) eV.

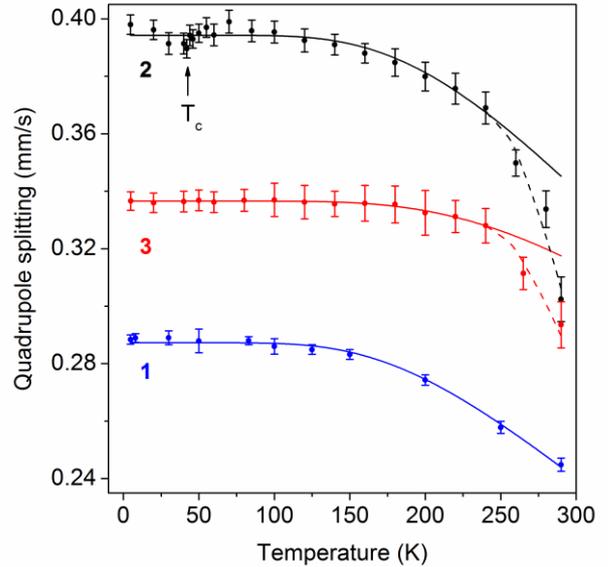

Fig. 5: (Colour on-line) Temperature dependences of the quadrupole splitting for **1** (blue), **2** (black) and **3** (red) and the fitting curves considering tetragonal distortion of FeSe$_4$ tetrahedra (eq. 2, solid line) and for the fluctuating EFG due to the thermoactivated motion of Li$^+$ ions (eq. 3, dashed line).

Another observation of the present work is a small decrease of $\Delta E_Q$ for **2** near $T_c$ as indicated in Fig. 5, quite small in magnitude and comparable to the error bars, but reaching its minimum exactly at $T_c$ = 42 K with 0.390(3) mm/s whilst being 0.396(4) mm/s at 100 K and 0.396(3) mm/s at 20 K.

This small decrease may be due to the appearance of magnetic fluctuations [27] and requires further investigations since the $\Delta E_Q$ vs. T dependence for **3** does not show any noticeable anomalies between 20 and 55 K, where the sample is not superconducting. Additionally, a slight decrease of the isomer shift for **2** can be found near $T_c$. This feature is not observed for **1** and **3** with much lower $T_c$ values. Hence, some variations of the hyperfine parameters of **2** in the vicinity of $T_c$ occur, but the nature of this phenomenon should be clarified in future.

**Conclusions.** – In summary, $^{57}$Fe-Mössbauer studies of superconducting FeSe and its derivatives reveal that $d$-electron density on $^{57}$Fe atoms increases upon intercalation with Li$^+$, NH$_2^-$ and NH$_3$ molecules between FeSe layers concomitant with a dramatic increase in $T_c$. The measurements also point to an enhanced value of the quadrupole splitting for the intercalate compound. To a large extent, this increase can be assigned to the Li+ ions, which provide an additional electric field gradient and the increase in electron density on Fe atoms resulting from the reduction of FeSe, i.e. "electron doping" by the Li$^+$ intercalation. Both $\delta$ and $\Delta E_Q$ enhancement can be considered as distinctive fingerprints of the high-$T_c$ superconductor Li$_x$(NH$_2$)$_y$(NH$_3$)$_{1-y}$Fe$_2$Se$_2$ (x=0.6; y=0.2). Despite increased density of $d$-electrons on Fe due to intercalation we cannot conclude that this is the only reason for the rise of $T_c$ from 8 K to 42 K. In addition, we observe small variations of the quadrupole splitting and isomer shift for **2** near $T_c$ that might be indicative of magnetic fluctuations connected with superconducting mechanism. Finally, we report on the observation of Li$^+$ motion in the intercalated compounds, which may be interesting materials exhibiting both high-$T_c$ superconductivity and Li$^+$ mobility.

\*\*\*

We thank the Deutsche Forschungsgemeinschaft (DFG) for financial support provided through Grant No. KS51/2-2 (priority program SPP-1458) and a research fellowship for S.J.S. (SE 2324/1-1). S.J.C. thanks the UK EPSRC for financial support (Grant EP/I017844). The authors are grateful to the group of Prof. R.J. Cava (Princeton University) for the synthesis of $\beta$-FeSe.